\documentclass[useAMS,usenatbib]{mn2e}

\usepackage[latin1]{inputenc}
\usepackage{graphicx}
\usepackage{subfigure}
\usepackage{rotate}
\usepackage{amsmath, float}
\usepackage{txfonts}
\usepackage{fleqn}
\usepackage{color}
\usepackage{comment}

\usepackage{mathrsfs}  % added packages
\usepackage{booktabs}
\usepackage{colortbl} 

\newcolumntype{G}{>{\columncolor[gray]{0.8}}l} % Gray column in tabular

\newcommand{\be}{\begin{equation}}
\newcommand{\ee}{\end{equation}}
\newcommand{\bdm}{\begin{displaymath}}
\newcommand{\edm}{\end{displaymath}}
\newcommand{\bea}{\begin{multline}}
\newcommand{\eea}{\end{multline}}
\newcommand{\ba}{\begin{align}}
\newcommand{\ea}{\end{align}}

\def\simlt{\mathrel{\hbox{\rlap{\hbox{\lower4pt\hbox{$\sim$}}}\hbox{$<$}}}}
\def\simgt{\mathrel{\hbox{\rlap{\hbox{\lower4pt\hbox{$\sim$}}}\hbox{$>$}}}}

\title[Polarization of synchrotron turbulent bubbles]
{Polarization properties of turbulent synchrotron bubbles: an approach
  based on   Chandrasekhar-Kendall functions}
\author[N. Bucciantini]{
N. Bucciantini$^{1,2,3}$\thanks{E-mail: niccolo@arcetri.astro.it}\\
$^{1}$INAF - Osservatorio Astrofisico di Arcetri, Largo E. Fermi 5,
I-50125 Firenze, Italy\\
$^{2}$Dipartimento di Fisica e Astronomia, Universit\`a degli Studi di Firenze, Via G. Sansone 1, 
I-50019 Sesto F.~no  (Firenze), Italy\\
$^{3}$INFN - Sezione di Firenze, Via G. Sansone 1, I-50019 Sesto F.~no  (Firenze), Italy}

\begin{document}
 
\date{Accepted / Received}

\maketitle

\label{firstpage}

\begin{abstract}
Synchrotron emitting bubbles arise when the outflow from a compact
relativistic engine, either a Black Hole or a Neutron Star, impacts on
the environment. The emission properties of synchrotron radiation
are widely used to infer the dynamical properties of these bubbles,
and  from them the injection conditions of the engine. Radio
polarization offers an important tool to investigate the level and
spectrum of turbulence, the magnetic field configuration, and possibly
the degree of mixing. Here we introduce a formalism based on
Chandrasekhar-Kendall functions that allows us to properly take into
account the geometry of the bubble, going beyond standard analysis
based on periodic cartesian domains. We investigate how different
turbulent spectra, magnetic helicity and particle distribution
function, impact on global properties that are easily accessible to
observations, even at low resolution, and we provide fitting formulae to
relate observed quantities to the underlying magnetic field structure. 

\end{abstract}

\begin{keywords}
radiation mechanisms: non-thermal - polarization - turbulence - radio
continuum: ISM - ISM: supernova remnants - ISM: bubbles
\end{keywords}

\section{Introduction}

Synchrotron emission originates from relativistic pairs, spiraling in
a magnetic field. In astrophysics synchrotron emission is a powerful
tool to investigate relativistic plasmas, and non-thermal particles
distributions. These are  the signatures of acceleration
processes related often to relativistic engines, like Neutron Stars
(as in the case of Pulsars) and Black Holes (as in the case of AGNs).

One of the key properties of synchrotron emission, is the high level
of linear polarization \citep{Westfold59a,Legg_Westfold68a}, that theoretically  can be as high as 70\%, for
the typical particles distribution functions that are observed (quite
often power-laws). It was
indeed thanks to its polarization, that synchrotron emission was
recognized for the first time in an astrophysical source, the Crab
nebula \citep{Baade56a,Oort_Walraven56a,Woltjer58a}.  When relativistic outflows interact with the
surrounding environment, they tend to form synchrotron emitting
bubbles. This happens for Pulsars inside Supernova Remnants \citep{Pacini_salvati73a,Rees_Gunn74a,Kennel_Coroniti84a,Gaensler_Slane06a,Bucciantini08a}, leading
to pulsar wind nebulae (often referred as {\it plerions}), that are
characterized by a broad-band emission ranging from Radio to X-ray
(and TeV due to Inverse Compton). It also happens for Radio Lobes
fed by the jet produced in AGNs \citep{Scheuer82a,Begelman_Blandford+84a,Begelman_Cioffi89a,Carilli_Perley+94a,Reich_Testori+01a}. The study of radio
emission and polarization enable us to infer
information on the strength and structure of the magnetic field, that
are relevant to understand the dynamics of these systems.  For
example, in pulsar wind nebulae, radio emission traces old particles,
that fill the entire volume of the nebula, and it is detectable also in old systems, where the injection from the pulsar
has faded away
\citep{Frail_Giancani+96a,Giancani_Dubner+97a,Roberts_Romani+99a,Ma_Ng+16a}. Recent
observations (e.g.  \citet{Ma_Ng+16a}), and a set of theoretical indications
based either on numerical simulations
\citep{Jun98a,Blondin_Chevalier+01a,Porth_Komissarov+14a} or on
spectral modeling 
\citep{Tang_Chevalier12a,Olmi_Del-Zanna+15a,Olmi_Del-Zanna+16a,Tanaka_Asano16a}
suggest that a non negligible level of turbulence should be present
in these systems. Some simplified model for the turbulent field has
already been presented attempting to explain some of the observed
results \citep{Ma_Ng+16a,Bucciantini_bandiera+17a}. The level of
turbulence in wind bubbles could be important to
understand how particles diffuse \citep{Tang_Chevalier12a}, the level of mixing and
penetration of the ambient medium \citep{Blondin_Chevalier+01a}, the
strength of the magnetic field 
(its ability to quench the cascade) and its role in the nebular
dynamics \citep{Bucciantini_Amato+04a}.

There is a vast literature that in the past years has focused on
modeling the polarized properties of synchrotron emission, in a
turbulent magnetic field, but most of it has focused on the galactic
background emission, associated to cosmic rays [e.g. \citet{Lazarian_Shutenkov90a,Lazarian_Chibisov91a,Waelkens_Schekochihin+09a,Junklewitz_Ensslin11a,Lazarina_Pogosyan12a,Lazarian_Pogosyan16a}]. Far less has
been done in the case of confined systems, like wind bubbles. Part of
the problem is due to the fact that the techniques developed for the
former are of little use in the latter. In general the approaches to
modeling the  polarized properties of synchrotron emission, in a
turbulent magnetic field, are based on the idea of an infinitely
extending volume, where the turbulence is homogenous
\citep{Thiebaut_Prunet+10a,Lazarian_Pogosyan15a,Zhang_Lazarian+16a,Herron_Buckhart+16a},
and make use of internal Faraday rotation, which is not present for a
pair plasma (as expected in pulsar wind nebulae). Assuming
homogeneous, infinitely extending turbulence, is
equivalent to neglecting any physical scale of the system
under investigation (they are all much larger that any of the scale of the
turbulent cascade). Unfortunately this is not the case for
wind bubbles, where the scales of the turbulent cascade are comparable
with the bubble dimension itself. On top of this we need to recall
that several mathematical tools have been developed through the years
to deal with turbulence in an homogenous infinite space, while the
case of a confined volume (where surface effects are important), being
more dependent on the specific boundary conditions, has been hardly
touched. Finally, numerical tools exist that are efficient, and easy
to implement to work in cartesian domains with periodic boundaries (think
the Fast Fourier Transform), that are not available on other domains
or for other geometries.

In this work we present a new approach to the study of turbulent
magnetic field configurations inside confined domains, and in
particular we select a spherical domain, which can be though of as a
good first-order approximation for an emitting bubble, based on an
alternative set of harmonic functions with respect to Fourier plane
waves. Our approach is based on Chandrasekhar Kendall functions
(CK) \citep{Chandrasekhar_Kendall57a,Chandra87a}. CK functions have
  been used for modeling spheromak configurations \citep{Vandas_Fisher+97a,Dasgupta_Janaki+02a}, plasmoid in the solar
corona \citep{Burglara88a,Chandra_Prasad91a,Farrugia_Osherovich+95a,Vandas_Fisher+91a}, and for dynamos in confined domains \citep{Mininni_Montgomery+06a,Mininni_Montgomery+07a,Brandenburg11a}. They
are the natural extension of Fourier modes to spherical geometry, but
they can be extended also to other geometry \citep{Rasband_Turner81a} as in the case of shells
\citep{Morse07a}, cylinders \citep{Marsh92a,Morse05a} or ellipsoids \citep{Ivanov_Kharshiladze85a}. They allow us to
set the proper boundary conditions for a confined field, and to take
into account the geometry of the emitting region in a formally correct
way (instead of truncating the emission at the edge of  a given volume without
care of the magnetic field structure). This allows us to take into
account volumetric effects.  Given that the scope of
this paper is to characterize the signature of a turbulent
field, providing estimates that could easily help guide the
interpretation of observations, we focus our attention on global
quantities that are easily measurable even at the typical low
resolution, at which synchrotron bubbles are observed.

In Sect.\ref{sec:ck}, we introduce the formalism that we adopt to model
a turbulent magnetic field confined within an emitting bubble. In
Sect.~\ref{sec:turb} we derive some simple scaling of quantities that
will be analyzed in Sect.~\ref{sec:res} where we present the result of
a statistical study of the global polarized properties for
synchrotron emitting bubbles. In Sect.~\ref{sec:conc} we present our conclusions.

\section{Magnetic field realization}
\label{sec:ck}

We are going to discuss here how to implement a magnetic field
realization on a spherical domain, such that we can enforce the specific
boundary condition of full confinement at the spherical surface bounding the domain
itself, in a rigorous manner. The idea is to use spherically adapted
harmonic function, instead of simple cartesian Fourier modes. Note that the
use of  adapted  harmonic functions can be extended to a
variety of different geometries, for which they are defined.

\subsection{Chandrasekar-Kendall functions}

In spherical coordinates $[r,\theta,\phi]$, given the scalar function:
\begin{align}
\Upsilon_{nlm}=J_l(k_{nl}r)Y_l^m(\theta,\phi),\label{eq:upsilon}
\end{align}
where $J_l$ is the spherical Bessel function of first kind of degree
$l$, $Y_l^m$ is the real spherical harmonic of degree $l,m$, and
$k_{nl}$ is the $n$-th zero of  $J_l$,  the vector field defined
by:
\begin{align}
\boldsymbol{B}_{\pm nlm} = \pm
k_{nl}\nabla\wedge[\Upsilon_{nlm}\boldsymbol{r}]+\nabla\wedge\nabla\wedge [\Upsilon_{nlm}\boldsymbol{r}]\label{eq:bdef}
\end{align}
is a force free field $\pm k_{nl}\boldsymbol{B}_{\pm nlm}
=\nabla\wedge \boldsymbol{B}_{\pm nlm}$, with vanishing radial
component of the unitary 2-sphere $\mathcal{S}$: $\boldsymbol{B}_{\pm
  nlm} \cdot \boldsymbol{e}_r =0$ at $r=1$. Being force free \citep{Woltjer58b,Molodensky74a}, this
vector field is also solenoidal, and it is known as {\it Chandrasekhar
  Kendall} functions. It can be shown that CK functions form a
complete basis for solenoidal vector
fields in the unitary ball $\mathcal{B}$, with vanishing component on
the boundary $\mathcal{S}$ [\citet{Yoshida92a,Mininni_Montgomery+06a,Mininni_Montgomery+07a}, see for example \citet{Deredtsov_Kazantsev+07a} for other basis]. CK functions can be written in terms of
real vector spherical harmonics:
\begin{align}
\boldsymbol{B}_{\pm nlm} &=& \frac{l(l+1)}{r}J_l(k_{ln}r)\boldsymbol{Y}_l^m +\left[\frac{{\rm
        d}}{{\rm
        d}r}J_l(k_{ln}r)+\frac{J_l(k_{ln}r)}{r}\right]\boldsymbol{\Psi}_l^m
  +\nonumber\\
  & &\pm k_{ln}
  J_l(k_{ln}r)\boldsymbol{\Phi}_l^m\label{eq:spherh}
\end{align}
where $\boldsymbol{Y}_l^m=Y_l^m \boldsymbol{e}_r$,
$\boldsymbol{\Psi}_l^m=r\nabla Y_l^m$ and $\boldsymbol{\Phi}_l^m  = r
\boldsymbol{e}_r\wedge\nabla Y_l^m$.
 CK functions are, moreover, orthogonal in the sense that:
\begin{align}
\int_{\mathcal{B}} \boldsymbol{B}_{anlm} \cdot \boldsymbol{B}_{efgh} d^3x =l(l+1)J^2_{l+1}(k_{nl})k_{nl}^2\delta^f_n\delta^g_l\delta^h_m\delta^e_a
\end{align}
where $a,e=\pm$ and $\delta^e_a=1$ only if $a$ and $e$ are the same
sign. 

\subsection{Magnetic field, power spectrum and CK functions}
Given the properties of CK functions any magnetic field bound to be
confined into a spherical region of unitary radius can be written as the sum of CK
modes:
\begin{align}
\boldsymbol{B}=\sum_{\pm}\sum_{n=1}^{\infty}\sum_{l=1}^{\infty}\sum_{m=-l}^{l}
c_{\pm nlm} \boldsymbol{B}_{\pm nlm}\label{eq:btdef}
\end{align}
where the first sum is done on CK functions of different sign. Being
the modes orthogonal, for the magnetic field defined in Eq.~\ref{eq:btdef}, it is
possible to write the total energy $E$,  the total helicity $H$, the
total current helicity $I$ in $\mathcal{B}$ as:
\begin{align}
E=\sum_{\pm}\sum_{n=1}^{\infty}\sum_{l=1}^{\infty}\sum_{m=-l}^{l}
c_{\pm nlm}^2 l(l+1)J^2_{l+1}(k_{nl})k_{nl}^2\label{eq:etdef]}\\
H=\sum_{\pm}\sum_{n=1}^{\infty}\sum_{l=1}^{\infty}\sum_{m=-l}^{l}
\pm c_{\pm nlm}^2 l(l+1)J^2_{l+1}(k_{nl})k_{nl}\label{eq:etdef]}\\
I=\sum_{\pm}\sum_{n=1}^{\infty}\sum_{l=1}^{\infty}\sum_{m=-l}^{l}
\mp c_{\pm nlm}^2 l(l+1)J^2_{l+1}(k_{nl})k_{nl}^3\label{eq:etdef]}
\end{align}
Note that modes with the same $n,l$ but different $m$ are degenerate
in energy. It is immediately obvious that the decomposition in CK functions,
allows one to control the helicity of the field and to build magnetic
field configurations having the same total energy but with various
possible prescription for the helicity:
\begin{itemize}
\item {\it maximal helicity realization} if $c_{-nlm}=0$ (or
equivalently  $c_{+nlm}=0$) for all $n,l,m$. In this case all CK modes
contribute an helicity of the same sign, and the total helicity is
maximal;
\item {\it zero helicity realization} if $c_{-nlm}=c_{+nlm}$ for all
  $n,l,m$. In this case the same power goes into CK modes of opposite
  helicity, and the total helicity vanishes;
\item {\it definite helicity realization} if $c_{-nlm}=0 \iff
  c_{+nlm}\ne0$ for all $n,l,m$. In this case all the power for each CK
  mode goes into a definite helicity state, even if the sign differs
  for different modes;
\item {\it random helicity realization} if no contraints are imposed
  on $c_{\pm nlm}$ for all $n,l,m$. In this case  the power for each CK
  mode is randomly distributed in the two possible helicity states. 
\end{itemize} 

In general, when using Fourier decomposition in finite cartesian domains, turbulent magnetic field realizations are buit assigning at
each mode of wave number $\boldsymbol{k}$ an amplitude such that when
integrating the energy contribution in the volume off all modes up to
any arbitrary value of the  wave number norm $k_{\rm max}$ one finds:
\begin{align}
E(k_{\rm max})=\sum_{\boldsymbol{k}}^{1<k<k_{\rm max}} E(\boldsymbol{k})
=\int_1^{k_{\rm max}} P(k)dk
\end{align}
where $E(\boldsymbol{k})$ is the energy associated with the Fourier
mode of wave number $\boldsymbol{k}$ and $P(k)$ is the on-shell power
spectrum (representing all the energy at a wave number $k$), and $k=1$
is the first allowed mode. Usually Fourier modes are quantized, given that in a finite
domain only modes that are periodic are allowed. Fourier decomposition
is however just a special case of decomposition into harmonic
functions, corresponding to plane waves, adapted to a cartesian domain. In different geometries other
harmonic decompositions might prove more suitable (one can for example
define properly the boundary
conditions on a spherical shell). In spherical
geometry, one can use for example spherical waves defined in term of
Bessel functions and spherical harmonics as Eq.~\ref{eq:upsilon}. It
can be shown that it is always possible to transform from one to the
other \citep{Mininni_Montgomery+06a,Mininni_Montgomery+07a,Liao_Su15a}. CK function can be seen as generalization to spherical geometry
of Fourier modes. In particular given that each CK mode has  maximal
helicity, they are the spherical equivalent of Beltrami waves. Boundary
conditions set the quantization of the modes [$ J_l(k_{nl})=0$]. 

If $P(k)\propto k^{-\delta}$, with $\delta >1$ then:
\begin{align}
\sum_{\pm}\sum_{n,l=1}^{k_{nl}<k_{\rm max}}\sum_{m=-l}^{l}
c_{\pm nlm}^2 l(l+1)J^2_{l+1}(k_{nl})k_{nl}^2\propto k_{\rm max}^{1-\delta}.
\end{align}
This condition translates into a condition on the coefficients of the
realization:
\begin{align}
\sum_{\pm}\sum_{m=-l}^{l}c_{\pm nlm}^2 = \frac{1}{l(l+1)}\frac{k_{nl}^{-\delta-1} }{J_{l+1}^2(k_{nl})}\label{eq:norm}
\end{align}
How this power is distributed among modes of different sign, depends
on the helicity prescription. On the other hand the distribution among
the various $m$ depends on the isotropy of turbulence. It is well
known that MHD turbulence is anisotropic \citep{Goldreich_Sridhar95a,Brandenburg_Lazarian13a}, with the anisotropy
dependent on scale. For simplicity, in our realizations we asume
isotropy. For a solenoidal magnetic field, isotropy implies that all
directions are equally probable for the orientation of each multipole
of degree $nl$. Given the known rotational properties of spherical
harmonics, in each realization we achive this by randomly distributing the power of the
multipole among the varios $m=-l,...,0,...,l$ subject to the
normalization given by Eq.~\ref{eq:norm}). In practice for each value
of $n$ and $l$ we generate
$2(2l+1)$ random number in the interval $[-1,1]$, (where the factor 2 comes from the presence of modes with two
sign: the $\pm$ part), we set each of the $c_{\pm nlm}$ equal to one
of these random numbers, and then  we renormalize them such that the
sum of their squared over $\pm$ and $m=-l,l$ satisfies the condition of  Eq.~\ref{eq:norm}. This ensures that: 1- on a
single realization, the spatial orientation of the various multipoles
$nl$, are uncorrelated; 2- that averaged
over many realizations, the orientation of each of the various multipoles,
over which the field is decomposed, is uniform on the 2-sphere. So,
fur us ``isotropy'' means that in each realization the various
multipoles are randomly oriented, and there is not preferential
alignement among them. Please not that the uniform assumption hold for the
ensemble of realizations.

\subsection{Numerical setup and simulated synchrotron maps}

Turbulent magnetic field realizations are built on a 3D cartesian grid
containing the unitary sphere. The resolution has been chosen such
that the
results on the quantities of interest are converged. Assuming a power law
distribution of emitting pairs:
\begin{equation}
n(\epsilon)= K \epsilon^{-(2\alpha+1)},
\label{eq:fdist}
\end{equation}
where $\epsilon$ is the energy in units $m_e c^2$, and $K$ in
principle could be a function of position. The emissivity toward the
observer (assumed to be located along the $x$-axis) at a frequency $\nu$ is:
\begin{equation}
j(\nu) = C\mid\boldsymbol{B}\times \boldsymbol{n}
\mid^{\alpha+1}\nu^{-\alpha}\,
\label{eq:j}
\end{equation}
where $\boldsymbol{n}=\boldsymbol{e}_x$, and $C$ is given by synchrotron theory
\begin{equation}
C=\frac{\sqrt{3}}{4}\frac{\alpha+5/3}{\alpha+1}\Gamma\left(\frac{\alpha
  +5/3}{2}\right)\Gamma\left(\frac{\alpha
  +1/3}{2}\right)\frac{e^3}{mc^2}\left(\frac{3e}{2\upi m c}
\right)^\alpha K\, \label{eq:cdef}
\end{equation}
Assuming the $y$ and $z$ coordinates to lay on the plane of the sky, it is possible to compute the maps of
the various Stokes parameters integrating the contribution of each
fluid element along the line of sight through the domain of the realization, according to
\begin{eqnarray}
I(\nu,y,z)&=&\int_{-\infty}^{\infty}j(\nu,x,y,z)\,{\rm d}x\,, \\
Q(\nu,y,z)&=&\frac{\alpha+1}{\alpha+5/3}\int_{-\infty}^{\infty}j(\nu,x,y,z)\cos{2\chi}\,{\rm d}x\,, \\
U(\nu,y,z)&=&\frac{\alpha+1}{\alpha+5/3}\int_{-\infty}^{\infty}j(\nu,x,y,z)\sin{2\chi}\,{\rm d}x\, ,
\end{eqnarray}
where  the local polarization position angle $\chi$ is the angle of
the emitted electric field vector in the plane of the
sky, such that
\begin{eqnarray}
\cos{2\chi} = \frac{B_y^2-B_z^2}{B_y^2+B_z^2}\,,\quad\quad\sin{2\chi}=-\frac{2B_yB_z}{B_y^2+B_z^2}\,.
\label{eq:stokesang2}
\end{eqnarray}
The total intensity $I$ and total polarized intensity
$I_p=\sqrt{U^2+Q^2}$, can be obtained integrating the various Stokes
parameters over the $(y,z)-$plane of the sky. 

\section{Cascade energy and total brightness}
\label{sec:turb}

If the turbulent cascade extends up to a maximum value $k_{\rm max}$
one can model the total emission from the unitary sphere of volume $V$ as the sum of
the emission coming from $n\approx Vk_{\rm max}^{3}$ regions inside
which the magnetic field can be taken as uniform. For a random
gaussian realization, in each of these regions, the field orientation
and strength will be randomly distributed. If the emitting particles
are uniformly distributed, then the total intensity $I$, integrated over
the emitting area, will just
be function of the magnetic field:
\begin{align}
I\propto \sum_{n} (B\sin{\theta})^{\alpha+1} = n
\langle(B\sin{\theta})^{\alpha+1}\rangle = n
\langle B^{\alpha+1}\rangle\langle(\sin{\theta})^{\alpha+1}\rangle 
\end{align}
where $\theta$ is the angle between the magnetic field $\boldsymbol{B}$ and the
direction of the observer $\boldsymbol{n}$,  we have used the definition of mean and the fact that direction
and strength of the magnetic field are uncorrelated variables, for
isotropic turbulence. For any quantity $Q$ having a Maxwellian
distribution (like the strength of the magnetic field), it can be
shown that $\langle Q^\alpha\rangle\propto\langle Q\rangle^{\alpha}$,
hence we find:
\begin{align}
I\propto \langle B^2\rangle^{(\alpha+1)/2} \propto E^{(\alpha+1)/2}
\end{align}
Then on can estimate how the total brightness depends on $k_{\rm
  max}$:
\begin{align}
I(k_{\rm max}) \propto E(k_{\rm max})^{(\alpha+1)/2}= [E(k_{\infty})
  -\delta E(k_{\rm max})]^{(\alpha+1)/2} \\
I(k_{\rm max}) =I(\infty)[1-\delta E(k_{\rm max})/  E(\infty)]^{(\alpha+1)/2}\label{eq:itrend}
\end{align}
where $E(\infty)$ and $I(\infty)$ are the magnetic energy and total
intensity of a realization extending to $k=\infty$
respectively, and we have introduced the quantity $\delta E(k_{\rm
max})$ which represents the difference of the magnetic energy between a
turbulent cascade extending to $k=\infty$ and one up to $k_{\rm max}$.
For the polarized intensity $I_{\rm p}=\sqrt{U^2+Q^2}$ instead, one must recall that unlike $I$,
$Q$ and $U$ are not positively defined. For a realization extending up
to $k_{\rm max}$, there will be $\approx k_{\rm max}^3$, domains that
contribute to the polarized intensity incoherently. So we expect that
 the contribution to the polarized intensity from these domains should scale as the variance of the total intensity. The difference
in polarized intensity between a
turbulent cascade extending to $k=\infty$ and one up to $k_{\rm max}$.
\begin{align}
\delta I_{\rm p}(k_{\rm max})\propto  \delta E(k_{\rm max})^{(\alpha+1)/2} k_{\rm
  max}^{-3/2}\propto  \delta E(k_{\rm max})^{(\alpha+1)/2+3/2(\delta-1)} \label{eq:iptrend}
\end{align}
suggesting that for typical values of $\delta$ and $\alpha$, the
polarized intensity should typically saturate within few percents
already at $\delta E(k_{\rm max}) \sim 0.2$.

\section{Results}
\label{sec:res}
We have computed models for various values of the turbulent spectral
index, $\delta = [4/3,3/2,5/3,11/6,2]$, and particle distribution
function $\alpha =[0,1/4,1/2]$. The turbulent index has been chosen in
order to include both Kolmogorov $\delta=5/3$ and Kraichnan $\delta
=3/2$ scaling. The particle distribution function, has been chosen
to target the typical radio spectrum of PWNe (for the Crab nebula
$\alpha = 1/4$). For each value of $\delta $ and $\alpha$ we computed
synchrotron maps for one hundred different random realizations, for
which the uniform assumptions holds as an ensemble, which
allowed us to compute the ensemble mean values of observable quantities and to
estimate their typical variances $\sigma$. In order to get converged
results on the quantities of interest, we found that each realization must include at least
$10^4$ CK modes (this means that we have to compute at least $10^4$
coefficnets $c_{\pm nlm}$ and the related Bessel functions and
spherical harmonics).  Taking the turbulent
spectrum to be a pure power-law, corresponds to the assumption that the
coherence scale of the turbulence is the size of the bubble itself.  We focus our attention to global
quantities that are easily accessible to simple measures. Quantities
like the two-points correlation function, or the Fourier spectrum of
the emission map, while far more informative, in general require the object to be observed at very
high resolution, which is often not the case for our typical targets (PWNe).

\subsection{Maximal helicity}

\begin{figure*}
	\centering
	\includegraphics[width=.95\textwidth,bb=40 10 490 250,clip]{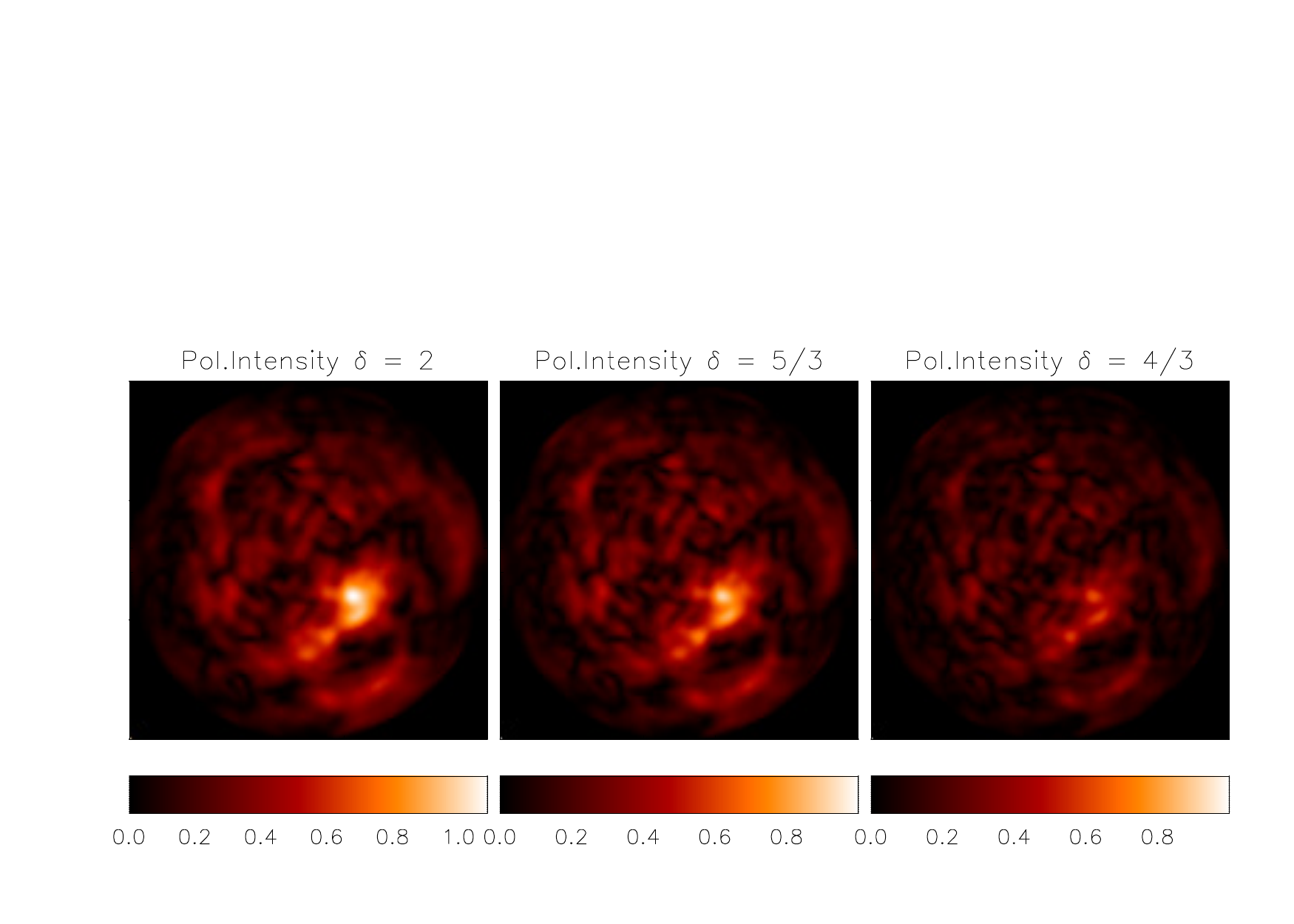}\\
	\caption{Polarized surface brightness maps, normalized to the
          mean total surface brightens, for the same magnetic
          realization (same orientations of the modes) but with
          various turbulent indexes, in the case of maximal
          helicity. The spectral index is set to $\alpha =1/4$
     }
	\label{fig:rmappe}
\end{figure*}

We present here the results in the simplest case  where the emitting
particles (pairs) are distributed uniformly in space ($K$ is constant
in space), and the magnetic field follows a
realization with maximal helicity: $c_{-nlm}=0$. In
Fig.~\ref{fig:rmappe} we compare brightness maps in polarized intensity
obtained from realizations with different turbulent power spectra but where the
power at each $k$ is distributed in the same proportion among the
degenerate modes (the relative orientation of the modes is the same),
such that the final maps have the same structure. As expected,
realizations with smaller values of $\delta$ tend to show
more fine structures, however the difference is quite small, and it would
require very high resolution and high signal-to-noise observations to
be detected. On the contrary we found that the polarized
intensity, once normalized to the average unpolarized brightness is
quite sensitive to the value of $\delta$. Another feature that is
commonly found in our realizations is the presence of one or two
highly polarized spots, where the typical polarized intensity  is
about twice the average nebular value, and which in general are not
coincident with the brightest regions in total intensity.  As shown in
the figure, these highly polarized spots, are not necessarily located
close to the center, first because synchrotron emissivity strongly
depends on the orientation of the magnetic field, and in general there
is no reason for the magnetic field to be preferentially orthogonal to
the line of sight close to the center of the bubble. Second being maps
in polarized intensity, one should remember that there are
depolarization effects once one integrates along the line-of-sight
over regions with different magnetic orientations. So polarized
emission will peak where, by chance of the realization, these
depolarization effects are minimal, and these regions might or might
not be close to the center. On the other hand the polarized fraction
tends to rise at the very edges of the bubble  where it can reach
values close to the theoretical maximum $(\alpha+1)/(\alpha+5/3)$,
in the body of the emitting bubble one finds either unpolarized
regions and regions with polarization as high as 50\% of the
theoretical maximum. Please note that the maps shown correspond just
to one realization, in order to provide an idea of the results to the
reader, but are not meant to represent any typical average. They are
are mean to illustrate how a different choice of the turbulent index
$\delta$, quenching or instrumental resolution affects the appearance
of map in polarized intensity.\\

\begin{figure}
	\centering
	\includegraphics[width=.50\textwidth]{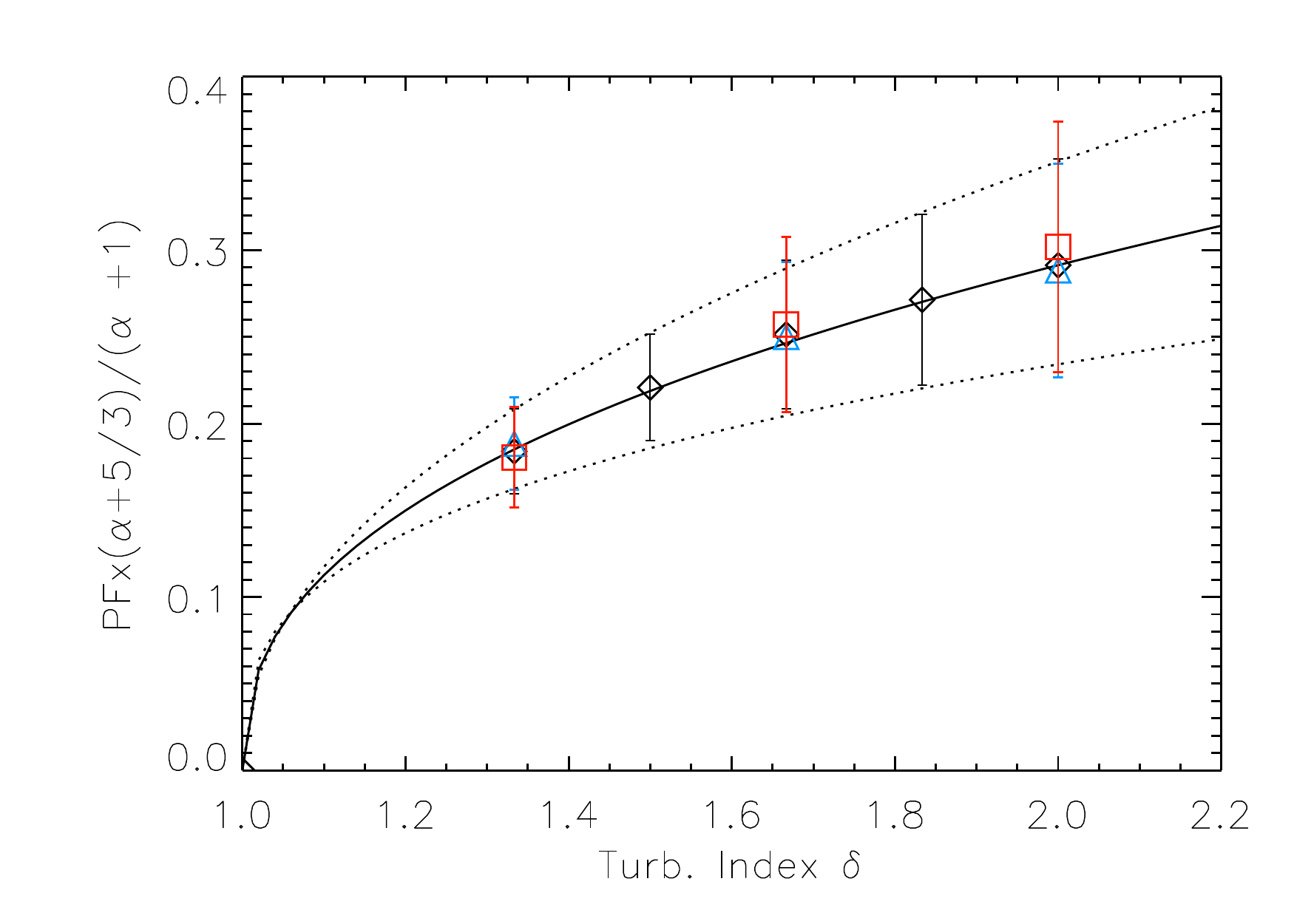}\\
	\caption{Integrated polarized fraction $PF$ normalized to the synchrotron factor
          $(\alpha+1)/(\alpha+5/3)$, as a function of the power spectral
          index of the turbulence $\delta$. Black diamonds are the
          mean values for $\alpha=1/4$, blue
          triangles for $\alpha =0$ , and red squares for
          $\alpha=1/2$. The error-bars represent the $2\sigma$ range. 
     }
	\label{fig:risultato1}
\end{figure}

In Fig.\ref{fig:risultato1} we show the mean (ensemble averaged over
many realizations) of the polarized fraction defined
as the ratio of total polarized intensity over total intensity: $
PF=I_p/I
$ with error-bars
indicating the $2\sigma$ range of the results computed over a hundred
different realizations. The $\delta=1$ point is
set at $PF=0$, according to the discussion of Sect.~\ref{sec:turb},
because for $\delta=1$ the total intensity is dominated by the
small scale components which carry most of the magnetic energy, hence the polarized fraction, which is instead
due to the large scale components, vanishes. 

%%%%%%%%%%%%%%%%%%%% Fig 1 %%%%%%%%%%%%%%%%%%%%%%%%%%%%%%%%%%%%%%
%%%%%%%%%%%%%%%%%%%%%%%%%%%%%%%%%%%%%%%%%%%%%%%%%%%%%%%%%%%%%%%%%

We see immediately that
the integrated polarized fraction shows a clear power-law dependence on
the turbulent index, while the dependence on the distribution
functions of the emitting pairs is, within the statistical
uncertainty of our realizations, simply given by the standard
synchrotron factor $(\alpha+1)/(\alpha+5/3)$. Indeed one can model our
result with the simple relation:
\begin{align}
{\rm Mean}[PF(\delta,\alpha)] = \frac{\alpha+1}{\alpha+5/3}[0.291(\delta
-1)^{0.413}]\label{eq:meanpf}
\end{align}
with the $2\sigma$ limits given by:
\begin{align}
[PF(\delta,\alpha)]^{+2\sigma} = \frac{\alpha+1}{\alpha+5/3}[0.351(\delta
-1)^{0.476}]\label{eq:sppf}\\
[PF(\delta,\alpha)]_{-2\sigma} = \frac{\alpha+1}{\alpha+5/3}[0.234(\delta
-1)^{0.333}]\label{eq:smpf}
\end{align}
Obviously these fits apply only to the range of $\delta$ we have
investigated. In particular, extrapolation to $\delta \rightarrow
\infty$ would lead to $PF \rightarrow \infty$ which is
unphysical. Indeed we find that, if only the mode $k=k_{11}$ is
present, then the mean $PF$ normalized to the spectral index factor is
$\sim 0.45$.
  
%%%%%%%%%%%%%%%%%%%% Fig 2 %%%%%%%%%%%%%%%%%%%%%%%%%%%%%%%%%%%%%%
%%%%%%%%%%%%%%%%%%%%%%%%%%%%%%%%%%%%%%%%%%%%%%%%%%%%%%%%%%%%%%%%%

\subsection{Zero helicity}

At the other extreme with respect to the maximal helicity case, there
are the zero helicity realizations with $c_{+nlm}=c_{-nlm}$. Attempts
to define observables able to constrain the level of the helicity of the
magnetic field, have been developed in the past, making use of
higher-order statistics and  rotation measure
\citep{Waelkens_Schekochihin+09a,Junklewitz_Ensslin11a}.
Here we focus on a simple estimate of the possible role of helicity
over integrated quantities. It can
easily be shown that for a magnetic field described as a simple linearly
polarized plane wave, the ratio of the polarized intensity over the
total intensity is a constant for every direction of the observer, and
in every position of the emitting domain, and the direction of
polarization is also constant. For a magnetic
field described as a circularly polarized plane wave, there are depolarization effects along the line
of sight (the direction of $\boldsymbol{B}\wedge \boldsymbol{n}$
changes throughout the emitting domain). So it is reasonable to expect
that a magnetic realization, with zero helicity, should show larger
polarized fractions. How large would depend on the power spectrum of
the turbulent realization. If the polarization properties of
the simulated bubble, are dominated by magnetic fluctuations of the
largest scale, to which only few modes of small $k$ contribute, then the dependence on the helicity will be
pronounced. If on the other hand polarization is dominated by
magnetic fluctuations on smaller scales, where many high $k$ modes
contribute incoherently, than one would expect to see only small
differences.

\begin{figure}
	\centering
	\includegraphics[width=.50\textwidth]{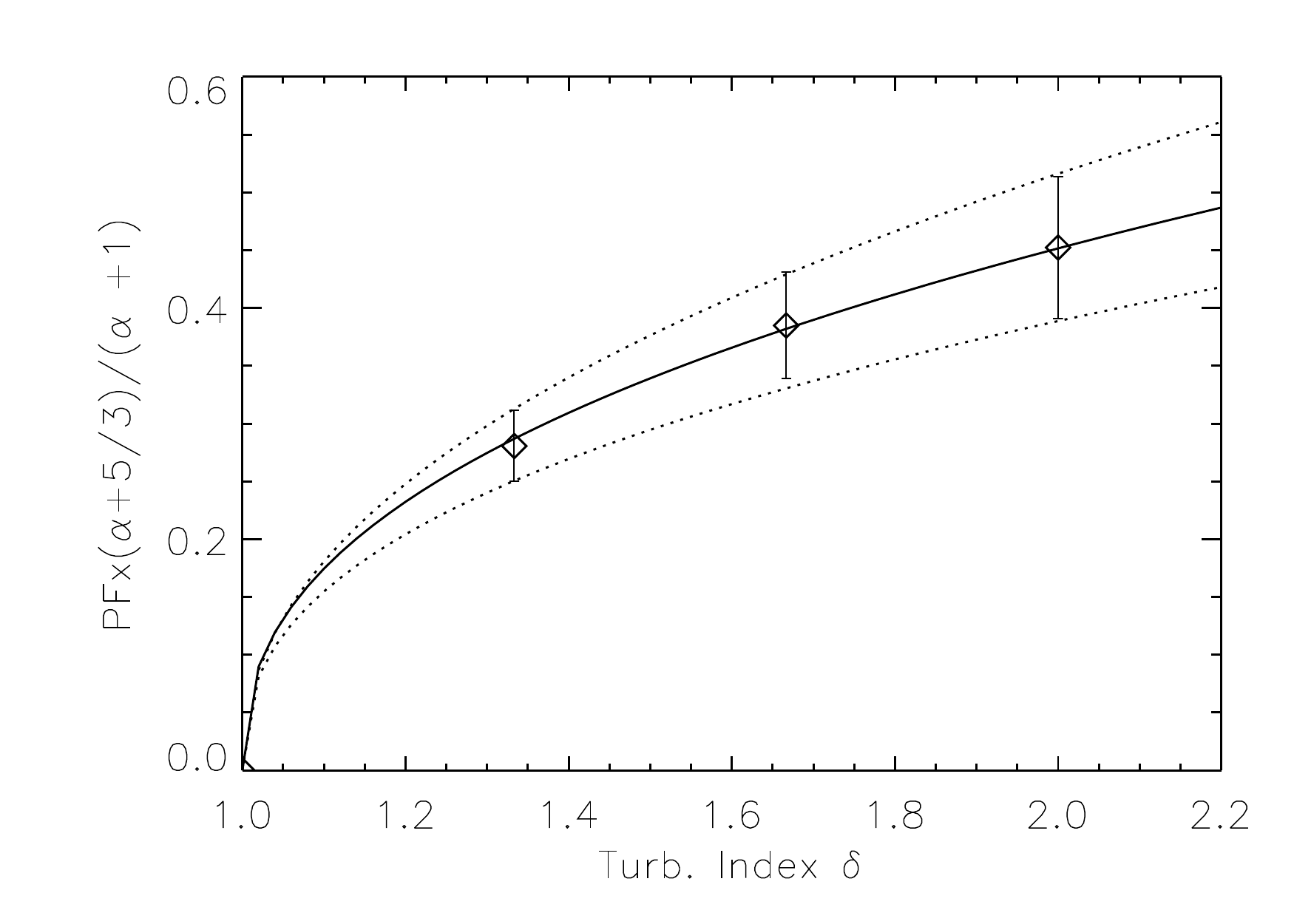}\\
	\caption{Integrated polarized fraction $PF$ normalized to the synchrotron factor
          $(\alpha+1)/(\alpha+5/3)$, as a function of the power spectral
          index of the turbulence $\delta$, for the zero helicity
          case: $H=0$. The spectral index is $\alpha=1/4$. The solid curve is a power-law fit to the
          averages, and is just given by Eq.~\ref{eq:meanpf}
          multiplied by a factor 1.52. The dotted curves represent
          power-law fit to the $2\sigma$ range.
     }
	\label{fig:risultatoh}
\end{figure}

Given the power spectrum we have adopted for our magnetic field
realizations, we found that the polarized fraction $PF$ is almost
completely set by the first few ($\sim 5$)
modes at small $k$, for all the values of $\delta$ we have simulated.
As a result the effects of a different helicity prescription are
substantial. On Fig.~\ref{fig:risultatoh} we show the same result of
Fig.~\ref{fig:risultato1}, but for $H=0$. Note that again we find that
the integrated polarized fraction can be described as a power-law
function of the turbulent index, but the typical values are 50\%
higher that the previous case and well beyond the $2\sigma$ range of
variance. Even the variance of the distribution is now $\sim 50\%$
larger. Despite the fact that the case $\delta =4/3$ has a flatter spectrum of
magnetic turbulence than the  $\delta=2$ case, we do not see any
appreciable, difference associated to the choice of helicity. This
suggests that, even at $\delta =4/3$, the large scales are still dominant.

%%%%%%%%%%%%%%%%%%%% Fig 3 %%%%%%%%%%%%%%%%%%%%%%%%%%%%%%%%%%%%%%
%%%%%%%%%%%%%%%%%%%%%%%%%%%%%%%%%%%%%%%%%%%%%%%%%%%%%%%%%%%%%%%%%

\subsection{Equipartition}

The magnetic field given by Eq.~\ref{eq:btdef} as the sum of
force-free modes, is itself not force free, and as such it is out of
dynamical equilibrium. In many situations of interest, like wind
bubbles, to which our approach is targeted, the energy of the magnetic
field is comparable to the  one in the emitting plasma. It is
common in these systems to assume equipartition between the particles
and the field, in order to infer physical informations from the
emission properties \citep{Pacholczyk70a}. The validity of such assumption is often
confirmed by more sophisticate models of the spectral energy
distribution. This requires the distribution of emitting particles to
be highly correlated to the magnetic field. In general however, when
dealing with possible variations of the density of emitting particles,
it is customary to assume a complete uncorrelation. In order to model a
distribution of emitting particles that follows an equipartition
recipe, we modified Eq.~\ref{eq:cdef} using a variable $K$ such that:
\begin{align}
K(\boldsymbol{r})\propto (B^2_{\rm max}-B^2(\boldsymbol{r}))
\end{align}
where $B(\boldsymbol{r})$ is the magnetic field strength at position
$\boldsymbol{r}$ and $B^2_{\rm max}$ is the maximum strength of the
magnetic field in the unitary ball $\mathcal{B}$ (which differs in
each realization). This ensures that the sum of the magnetic
and particle energy is constant throughout the emitting volume.

Interestingly we find that the mean polarized fraction and its
variance are unchanged with respect to the case of a uniform electron
distribution, even if the surface brightness in the emission maps,
appears shallower. Given that emission scales linearly with the
particle number density, one can safely conclude that for an electron
distribution given as:
\begin{align}
K(\boldsymbol{r})= K_o + C(B^2_{\rm max}-B^2(\boldsymbol{r}))
\end{align}  
the polarized fraction is independent of the value of $C$.

\subsection{Cascade quenching and instrumental resolution}

\begin{figure*}
	\centering
	\includegraphics[width=.95\textwidth,bb=40 10 490 250,clip]{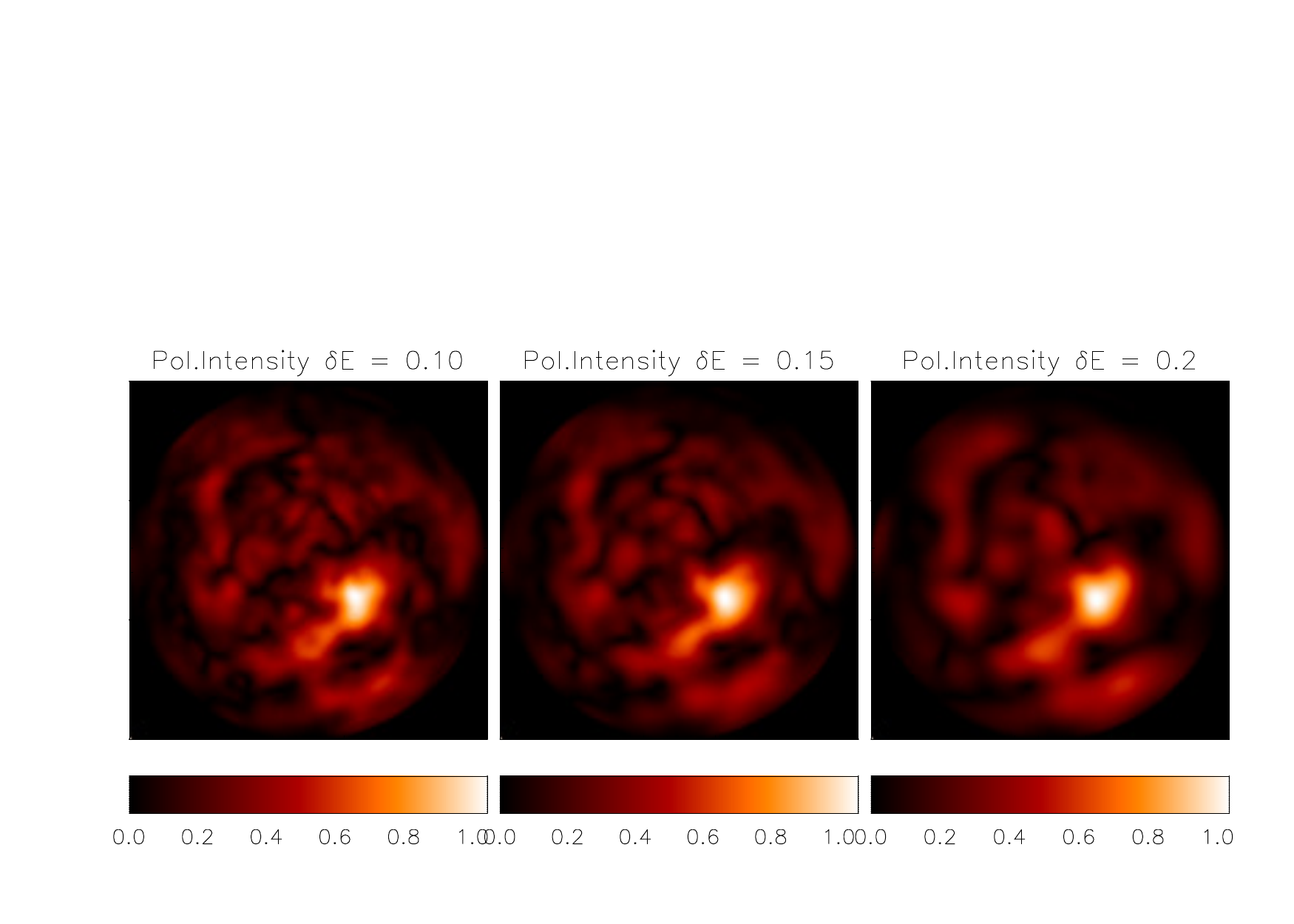}\\
	\caption{Polarized surface brightness maps, normalized to the
          mean total surface brightens, for the same magnetic
          realization (with $\delta=2$ and $\alpha =1/4$)  in the case of maximal
          helicity, but for turbulent magnetic cascades truncated at
          varios valued of $\delta E(k_{\rm max})$.
     }
	\label{fig:qmappe}
\end{figure*}

\begin{figure}
	\centering
	\includegraphics[width=.45\textwidth,bb=40 10 490 350,clip]{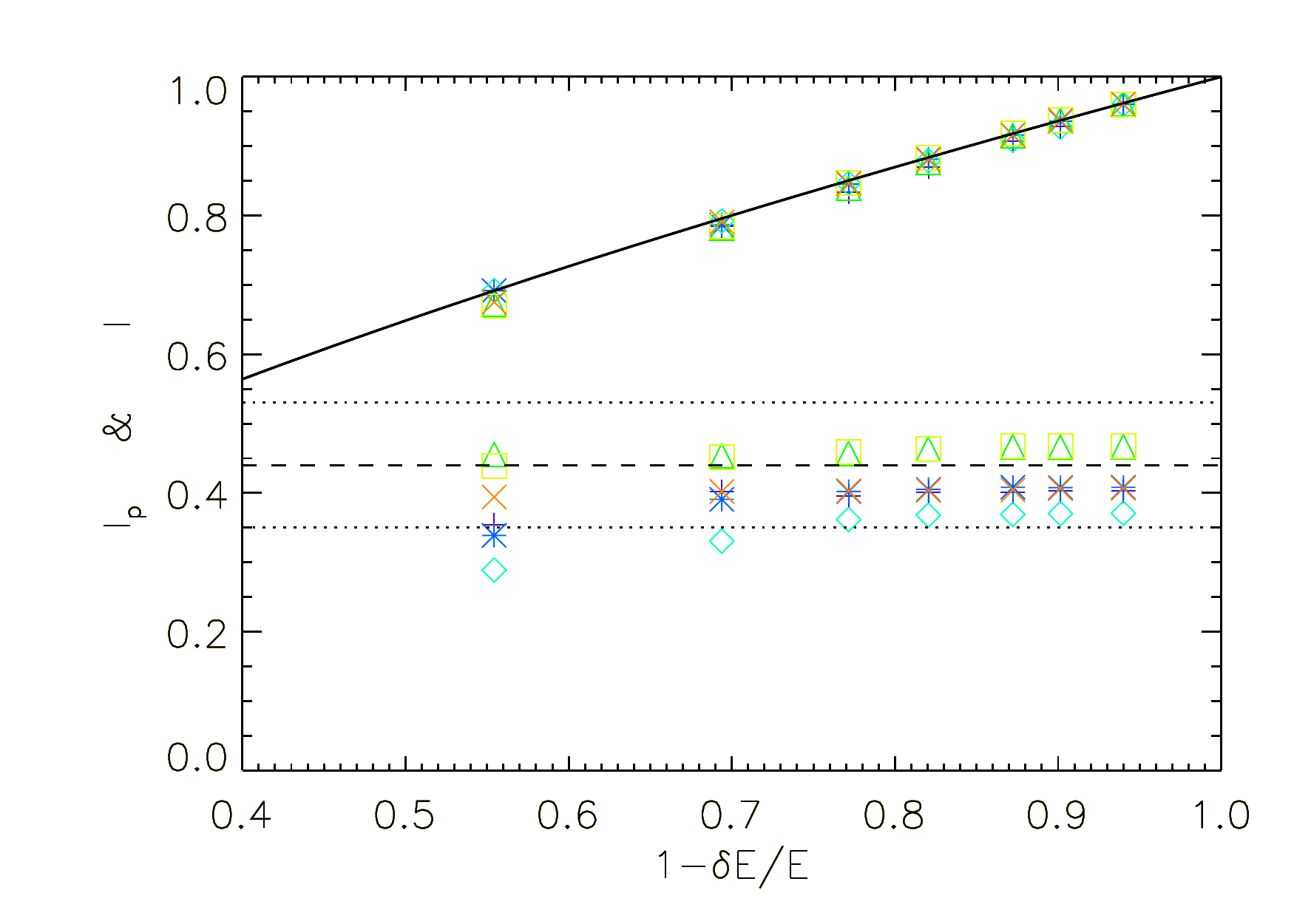}\\
	\caption{Trend of the total intensity and polarized intensity,
          as a function of cascade quenching, for the maximal helicity
          case $\delta =2$
          and $\alpha =1/4$. The upper points correspond to the total intensity
          $I$ of several realizations as a function of $\delta
          E(k_{\rm max})/E(\infty)$ (each point-style/color selects a
          realization set with a given configuration of the modes,
          progressively truncated at higher $k$). The lower points
          refer to the corresponding polarized intensity $I_p$. $I$
          and $I_p$ are normalized to the total intensity in the limit
          $k\rightarrow\infty$. The solid line represents the function
          $[1-(1-\delta E(k_{\rm max}))^{(\alpha+1)/2})]$, that provides a
          good fit to the trend of the intensity. The dashed line
          represents the mean polarized fraction as given by
          Eq. \ref{eq:meanpf}, and the dotted lines the $2\sigma$
          range as given by Eq.~\ref{eq:smpf}-\ref{eq:sppf}, for the
          limit $k_{\rm max}\rightarrow\infty$. 
     }
	\label{fig:quench}
\end{figure}

The results discussed previously, assume that the turbulent cascade
extends all the way to $k\rightarrow\infty$, and that maps have infinite
spatial resolution. We can however take into account both the effect
of turbulent quenching (if the cascade is truncated at a given $k_{\rm
max}$) and of instrumental resolution. In Fig.~\ref{fig:qmappe} we
show how the map of polarized intensity changes depending where the
cascade is truncated. In Fig.~\ref{fig:quench} we plot the total
intensity and polarized intensity, for some of our realizations,
showing how they change
depending on the truncation of the power-law distribution of the magnetic field. In
particular the cutoff is parametrized in terms of the {\it relative cascade
  energy}: $1-\delta E(k_{\rm max})/E(\infty)$.

We find that in general the polarized intensity saturates already for
values of $\delta E(k_{\rm max}) \sim 0.3 E(\infty)$, coherently with our estimate Eq.~\ref{eq:iptrend}, on the other hand the total
intensity keeps increasing quasi-linearly according to
Eq.~\ref{eq:itrend}. This, as discussed, reflects tha fact that
while the total intensity is always a positively defined quantity,
and so it increases with the addition of small scale modes, the polarized intensity is derived from Stokes
parameters $U$ and $Q$ which are not positively defined, and so undergo
cancellation effects. Interestingly we find that, for $E(k_{\rm max})
> 0.6$ the following relation for the
polarized fraction holds with a few percent accuracy:
\begin{align}
PF(E(k_{\rm max})) =[1-(1-\delta E(k_{\rm max}))^{(\alpha+1)/2})]^{-1} PF(E(\infty))
\end{align}
where $PF(E(\infty))$  corresponds to the polarized fraction of a cascade
extending to $k=\infty$ and whose mean
and variance are given in Eq.s~\ref{eq:meanpf}-\ref{eq:smpf}. 

%%%%%%%%%%%%%%%%%%%% Fig 4 %%%%%%%%%%%%%%%%%%%%%%%%%%%%%%%%%%%%%%
%%%%%%%%%%%%%%%%%%%%%%%%%%%%%%%%%%%%%%%%%%%%%%%%%%%%%%%%%%%%%%%%%

%%%%%%%%%%%%%%%%%%%% Fig 5 %%%%%%%%%%%%%%%%%%%%%%%%%%%%%%%%%%%%%%
%%%%%%%%%%%%%%%%%%%%%%%%%%%%%%%%%%%%%%%%%%%%%%%%%%%%%%%%%%%%%%%%%

%%%%%%%%%%%%%%%%%%%% Fig 4 %%%%%%%%%%%%%%%%%%%%%%%%%%%%%%%%%%%%%%
\begin{figure*}
	\centering
	\includegraphics[width=.95\textwidth,bb=40 10 490 250,clip]{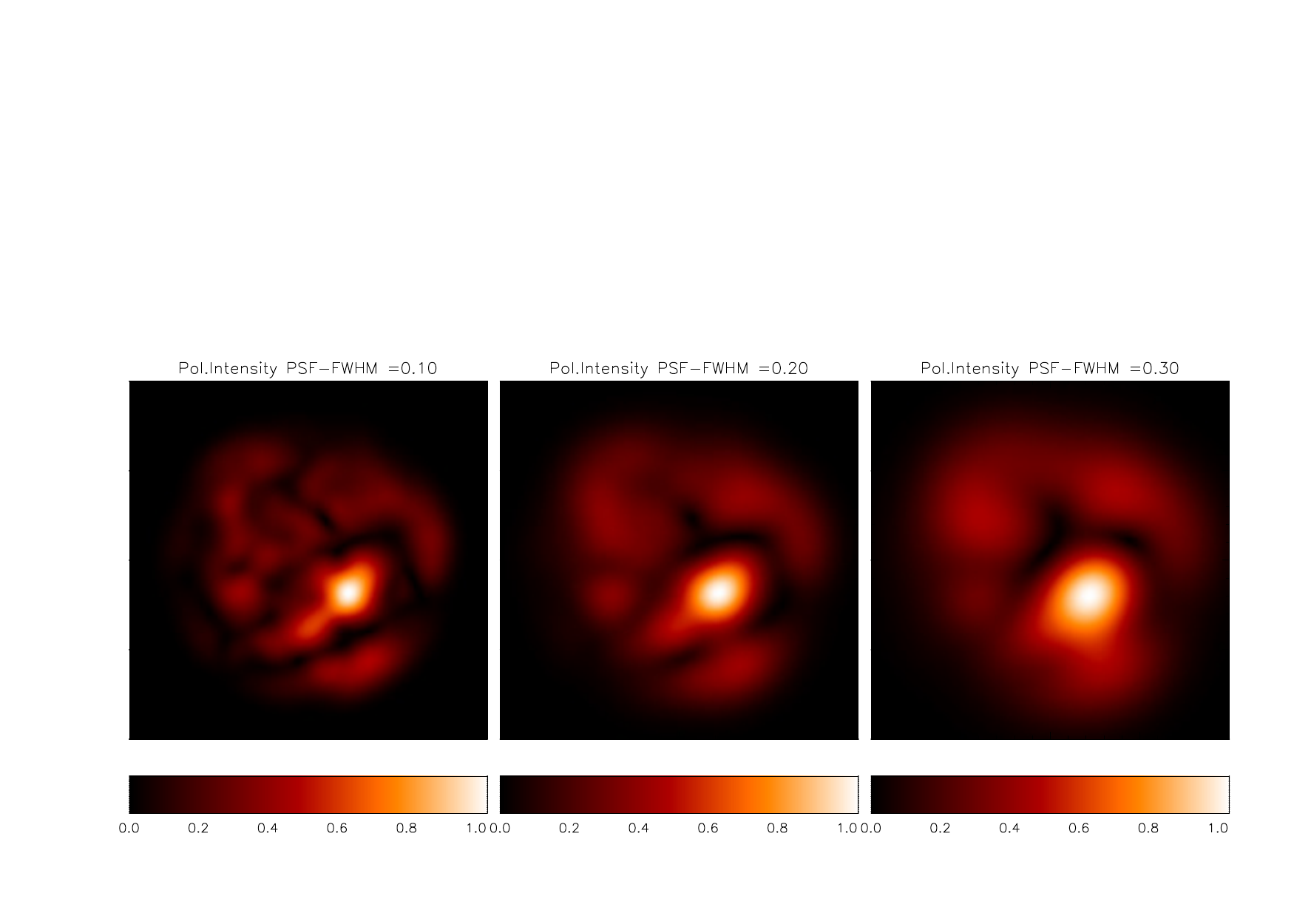}\\
	\caption{Polarized surface brightness maps, normalized to the
          mean total surface brightens, for the same magnetic
          realization (with $\delta=2$ and $\alpha =1/4$)  in the case of maximal
          helicity, but for different full-width half maximum
          (FWHM) of the point spread function. From left to right: $0.10$, $0.20$ and $0.30$ times
          the diameter of the bubble.
     }
	\label{fig:resol1}
\end{figure*}
%%%%%%%%%%%%%%%%%%%%%%%%%%%%%%%%%%%%%%%%%%%%%%%%%%%%%%%%%%%%%%%%%

%%%%%%%%%%%%%%%%%%%% Fig 5 %%%%%%%%%%%%%%%%%%%%%%%%%%%%%%%%%%%%%%
\begin{figure}
	\centering
	\includegraphics[width=.45\textwidth,bb=10 10 490 350,clip]{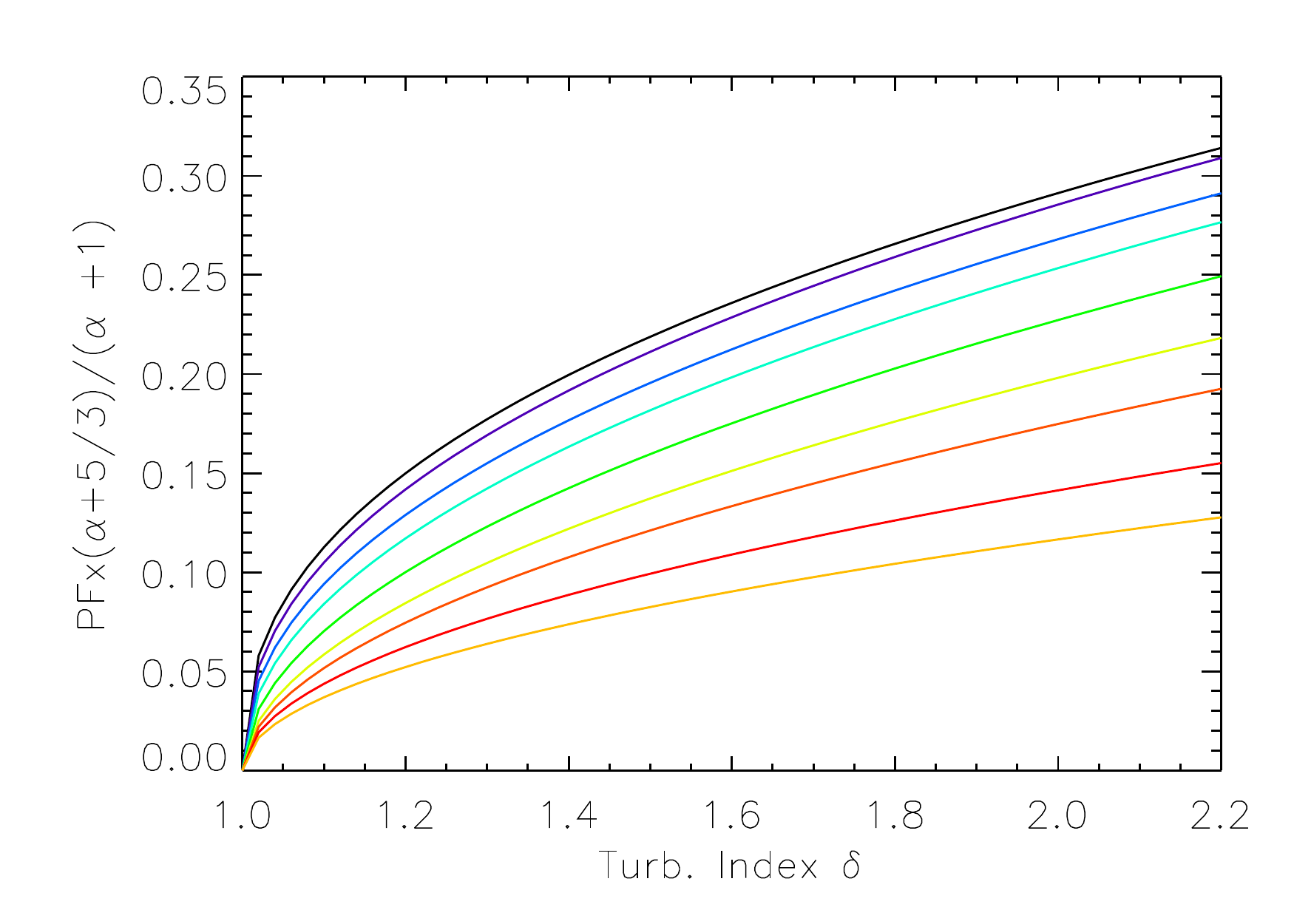}\\
	\caption{Best fits of the mean of the integrated polarized
          fraction PF normalized to the synchrotron factor $(\alpha +
          1)/(\alpha + 5/3)$, as a function of the power spectral index
          of the turbulence $\delta$. The spectral index is $\alpha
          =1/4$. The curves refer to different instrumental
          resolution. From top to bottom the full-width half-maximum
          of the point spread function
          is $[0,0.02,0.04,0.06,0.10,0.15,0.20,0.30,0.40]$ times the
          diameter of the emitting bubble.
     }
	\label{fig:resol2}
\end{figure}
%%%%%%%%%%%%%%%%%%%%%%%%%%%%%%%%%%%%%%%%%%%%%%%%%%%%%%%%%%%%%%%%%

On the other hand, instrumental resolution acts to decrease the level of
polarization, because, while not altering the total intensity, it
introduces cancellation effects of the Stoke's parmantes $U$ and
$Q$. In Fig.~\ref{fig:resol1} we show how the  polarized
emission maps change due to instrumental resolution assuming a
gaussian point spread function of varying width. Interestingly by
comparing the first panel in Fig.~\ref{fig:resol1} with the last one
in Fig.~\ref{fig:qmappe}, we see that on the morphology of the
polarized emission, the effect of a truncated magnetic cascade is
analogous to the one due to instrumental resolution. The two maps
are hardly distinguishable, especially with reference to the brightest
parts. What really distinguishes the two cases, is the net polarized
fraction, because quenching and instrumental resolution, act
differently on the total intensity. In
Fig.~\ref{fig:resol2}, we show how the mean polarized fraction of our realizations
changes, depending on instrumental resolution and for different values
of $\delta$. Interestingly we find that even in this case, we can fit
the dependence of the mean value on $\delta$ with a power law. The variance instead shows only minor
changes (at most $\sim 20$\%) with respect to the case of infinite
resolution. 

By comparing Fig.~\ref{fig:rmappe} with Fig.~\ref{fig:resol1}, or Fig.~\ref{fig:qmappe}, we see that
the structure of the polarized emission changes. While for an infinite
cascade, at very high resolution, the polarized intensity, show many
small scale features, in the other cases, these do not smear out into
a more uniform map, but ends up merging into larger structures that
have the general shape of loops. Interestingly a similar pattern was
seen in the body of G327.1-1.1 \citep{Ma_Ng+16a}. It was also observed that the
magnetic field direction seems to follow the loops. In that paper a
model was put forward to explain the level of polarization and the
typical scales observed, however due to the simplicity of the approach
it was not possible to reproduce the loopy structure of the magnetic
field, and the correlation between direction and bright features. The
authors suggested that a possible reason for this discrepancy was to be
looked for in the fact that the model did not enforce the solenoidal
condition on the magnetic field. In our model where the solenoidal
condition is enforced by construction, we indeed recover exactly this
king of behavior, with the direction of the magnetic field inferred
from Stoke's parameter aligned with the bright features. 

\subsection{Two points correlation}

\begin{figure}
	\centering
	\includegraphics[width=.45\textwidth]{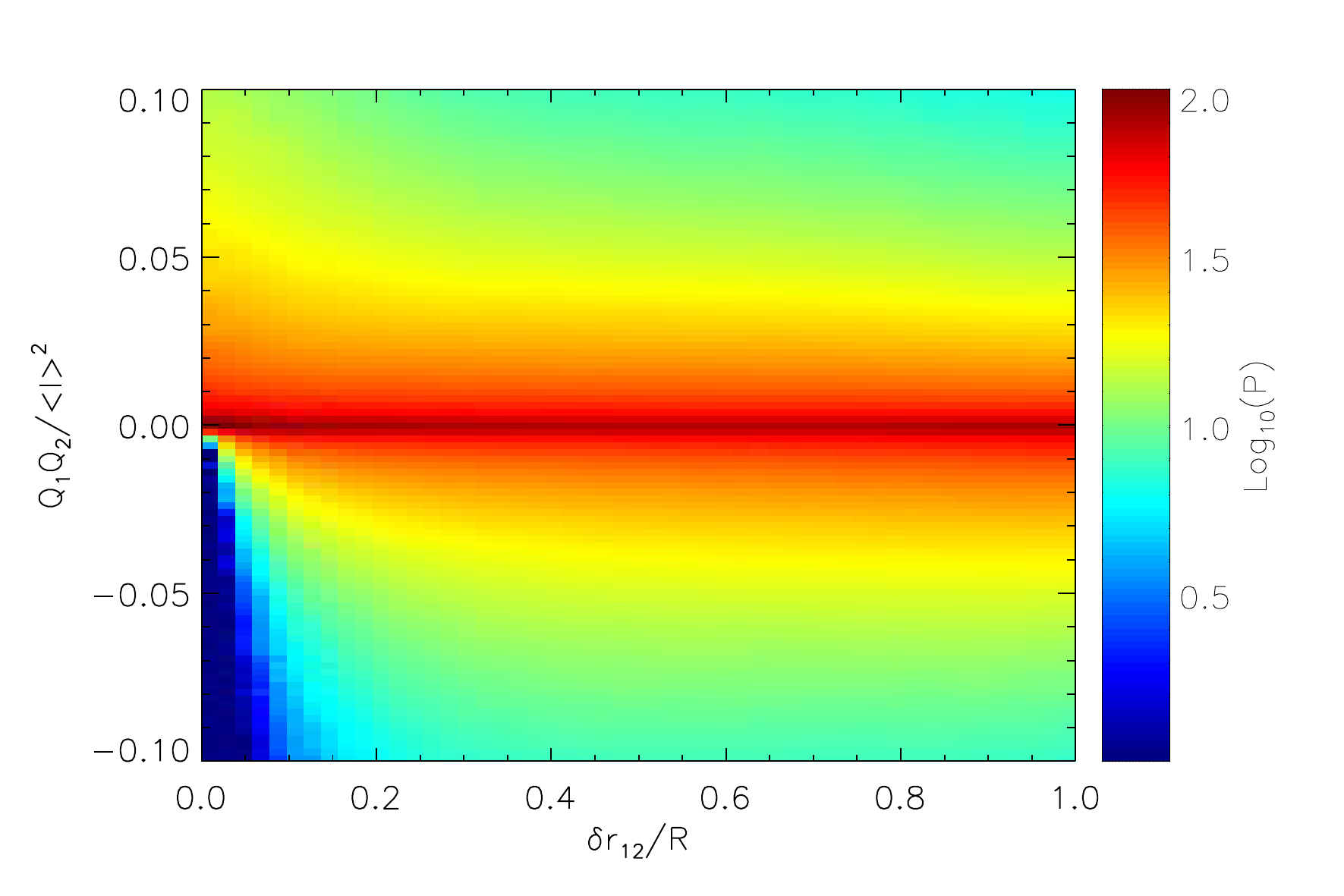}\\
        \includegraphics[width=.45\textwidth]{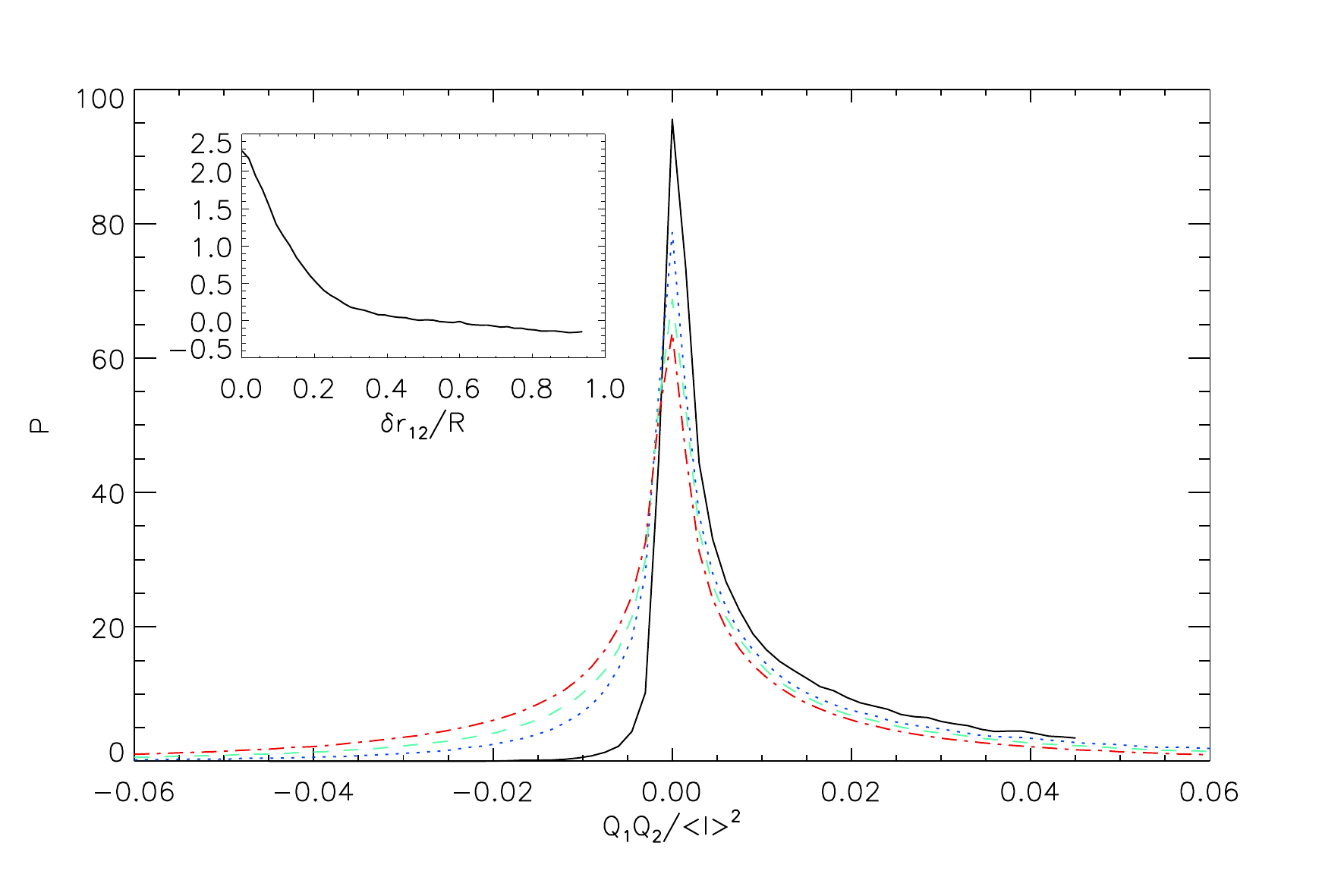}\\
	\caption{Upper panel: probability distribution function of the
          two-point correlation $Q_1Q_2(\delta r_{12})$ normalized to
          the mean squared total surface brightness  $\langle I
          \rangle^2$, as a function of the separation between the two
          points $\delta r_{12}$. Lower panel:  probability distribution function of the
          two-point correlation $Q_1Q_2(\delta r_{12})$ for selected
          values of $\delta r_{12}$: 0.02 solid-black line, 0.1
          dotted-blue line, 0.2 dashed-green line, 0.5 dash-dotted-red
          line. The insert shows the average value of  $Q_1Q_2(\delta
          r_{12})/ \langle I\rangle^2$ (multiplied times 100), as a function of $\delta r_{12}$.
     }
	\label{fig:2pt1}
\end{figure}

In the previous sections we have investigated the trends of global
integrated quantities, like the total polarized flux and fraction,
which are easily accessible even to simple observations. Here, in
order to provide a quantitative estimate of the  small scale
properties of our simulated maps, we investigate the two-points
correlation function of the Stoke's parameters defined (in the case of
$Q$ and analogously for $U$), as:
\begin{align}
Q_1Q_2(\delta r_{12})=Q(\boldsymbol{r}) Q(\boldsymbol{r}+\delta\boldsymbol{r})
\end{align}
where $1$ and $2$ label the two points located respectively at
$\boldsymbol{r}$ and $\boldsymbol{r}+\delta\boldsymbol{r}$, and
$\delta r_{12}=|\delta\boldsymbol{r} |$.
In the top panel of Fig.~\ref{fig:2pt1} we show, for a Komogorov distribution $\delta =5/3$, the probability distribution $P$ of the
value $Q_1Q_2(\delta r_{12})$, normalized for convenience to the mean total
surface brightness squared $\langle I \rangle^2$ as a function of the
separation $\delta r_{12}$.  It is evident the skewness of the
distribution for $\delta r_{12} <0.2 R$, where the two point
correlation function is dominated by positive values suggesting that
the polarized features are correlated, while for $\delta r_{12} >0.4
R$, the distribution is symmetric pointing to uncorrelated
features. This is made more evident in the bottom panel of
Fig.~\ref{fig:2pt1}, where the probability distribution is shown for
various values of  $\delta r_{12}$, and where the mean value of the
two point correlation function is also shown. 

%%%%%%%%%%%%%%%%%%%% Fig 5 %%%%%%%%%%%%%%%%%%%%%%%%%%%%%%%%%%%%%%
%%%%%%%%%%%%%%%%%%%%%%%%%%%%%%%%%%%%%%%%%%%%%%%%%%%%%%%%%%%%%%%%%

\section{Conclusions}
\label{sec:conc}

In this work we have introduced a novel approach to the study of
synchrotron emission from a turbulent magnetic field realization, based
on a set of harmonic functions (the Chandrasekhar-Kendall functions)
that allows us to take into account the geometrical properties of the
source, and to enforce correct surface-bounday conditions, going
beyond the standard small scale approximation, typical of many
previous studies. This is particularly relevant for synchrotron
emitting bubbles, where current instrumental resolution does not allow
us to sample structures that are much smaller than the bubble size
itself. The approach offers moreover the possibility to take into
account anisotropy in the turbulent cascade, and to control the level and distribution of
helicity among the modes.

We try to investigare how global properties, like the integrated
polarized fraction, scale depending on the spectrum of turbulence,
helicity distribution, spectral index of the emitted radiation,
resolution and cascade quenching. Interestingly we found that the
spatial distribution of the emitting particles (either uniform or
anticorrelated with the magnetic field strength) does not affect such
global properties. While the dependence on the spectral index (tied to
the particle distribution function) follows the standard synchrotron
law. We found that the mean polarized fraction of our realizations
can be fitted with a power-law with respect to the turbulent index
$\delta$ and that this trend holds also for different helicity and
instrumental resolution.

To illustrate an application of our results, let us consider for example G327.1-1.1 \citep{Ma_Ng+16a}. This is
middle age Pulsar Wind Nebula, with the pulsar outside the main radio
body of the nebula. In radio a loopy structure is observed, with the
magnetic field inferred from linear polarization aligned with the
bright features. The spectral index in radio is $\alpha=0.3$. The
body observed at 3cm and 6cm, with a resolution corresponding to a
FWHM $\sim 0.05$ times the nebular diameter, shows an average polarized fraction $\sim
15-20\%$, with a single highly polarized spot $\sim 30\%$ at the
center, and an increasing polarization toward the edges, mostly an
artifact due to the low  local surface brightness. With reference to
Fig.\ref{fig:resol2} we find that the lower value coincides with the
expected mean of PF in the Kolmogorov case $\delta =5/3$, with maximal helicity, while the
upper limit, higher than the $2\sigma$ range, would suggest some
cascade quenching with $\delta E \sim 0.7E(\infty)$. Obviuolsy, taking
into account the typical variance of our realizations, the detected
polarized fraction, is compatible with turbulent spectra in the range
$1.5<\delta<2.1$, and the upper limit cound be due to a different
helicity. It is howerver interesting that our model allows us, to set
some constraint on the tubulence inside such system: the preferred
model suggests a Kolmogorov turbulence, either with low helicity or
some quenching.

\section*{Acknowledgements}
The author acknowledge support from the PRIN-MIUR project prot. 2015L5EE2Y \emph{Multi-scale simulations of high-energy astrophysical plasmas}.

\bibliography{Bib}{}
\bibliographystyle{mn2e}

\end{document}